\documentclass[onecolumn,showpacs]{revtex4}

\usepackage{graphicx}
\usepackage{dcolumn}
\usepackage{bm}

\input epsf

\begin{document}

\title{\Large Non-adiabatic collapse of a quasi-spherical radiating star}

\author{\bf Ujjal Debnath}
\email{ujjaldebnath@yahoo.com}
\author{\bf Soma Nath}
\author{\bf Subenoy Chakraborty}
\email{subenoyc@yahoo.co.in}

\affiliation{Department of Mathematics, Jadavpur University,
Calcutta-32, India.}

\date{\today}

\begin{abstract}
A model is proposed of a collapsing quasi-spherical radiating
star with matter content as shear-free isotropic fluid undergoing
radial heat-flow with outgoing radiation. To describe the
radiation of the system, we have considered both plane symmetric
and spherical Vaidya solutions. Physical conditions and
thermodynamical relations are studied using local conservation of
momentum and surface red-shift. We have found that for existence
of radiation on the boundary, pressure on the boundary is not
necessary.
\end{abstract}


\maketitle

\section{\normalsize\bf{Introduction}}

In Einstein gravity, gravitational collapse with realistic
astronomical matter distribution is an important problem of
astrophysics. Usually, the formation of compact stellar objects
such as white dwarf and neutron star are preceded by a period of
radiative collapse. Hence for astrophysical collapse, it is
necessary to describe the appropriate geometry of interior and
exterior regions and to determine proper junction conditions
which allow the matching of these regions.\\

 The study of gravitational collapse was started long ago in 1939 by
Oppenheimer and Snyder [1]. They studied dust collapse with a
static Schwarzschild exterior while interior space-time is
described by Friedman like solution. Since then several authors
have extended the above study of collapse of which important and
realistic generalizations are the following: (i) the static
exterior was studied by Misner and Sharp [2] for a perfect fluid
in the interior, (ii) using the idea of outgoing radiation of the
collapsing body by Vaidya [3], Santos and collaborations [4-9]
included the dissipation in the source by allowing radial heat
flow (while the body undergoes radiating collapse). Recently,
Ghosh and Deskar [10] have considered collapse of a radiating star
with a plane symmetric boundary (which has a close resemblance
with spherical symmetry [11]) and have concluded with some general
remarks.\\

So far most of the studies have considered radiating star with
interior geometry as spherical. But in the real astrophysical
situation the geometry of the interior of a star may not be
exactly spherical, rather quasi-spherical in form. So it will be
interesting to study quasi-spherical interior geometry of a
radiating star. In this paper,we have considered the interior
space-time $V^{-}$ by Szekeres' model [12, 13] while for exterior
geometry $V^{+}$ we have considered plane symmetric Vaidya
space-time. The plan of the paper is as follows: Exact heat flux
solution in Szekeres' model has been presented in section II. The
junction conditions, physical properties and thermodynamical
relations are shown in section III. The paper ends with a
discussion in section IV.

\section{\normalsize\bf{Solution in Szekeres model with heat flux}}
The space-time metric for szekeres' model in ($n+2$) dimension is
in the form [13, 14]
\begin{equation}
ds^{2}_{-}=-dt^{2}+e^{2\alpha}dr^{2}+e^{2\beta}\sum^{n}_{i=1}dx_{i}^{2}
\end{equation}

where $\alpha$ and $\beta$ are functions of all the ($n+2$)
space-time variables. The stress-energy tensor of a non-viscous
heat conducting fluid has the expression
\begin{equation}
T_{\mu\nu}=(\rho+p)u_{\mu}u_{\nu}+pg_{\mu\nu}+q_{\mu}u_{\nu}+q_{\nu}u_{\mu}
\end{equation}
where $\rho,~ p, ~q_{\mu}$ are the fluid density, isotropic
pressure and heat flow vector. We take the heat flow vector
$q_{\mu}$ to be orthogonal to the velocity vector i.e.,
$q_{\mu}u^{\mu}=0$. For comoving co-ordinate system we choose
$u^{\mu}=(1,0,0,0,...,0)$ and
$q^{\mu}=(0,q,q_{1},q_{2},...,q_{n})$ where
$q=q(t,r,x_{1},...,x_{n})$ and
$q_{i}=q_{i}(t,r,x_{1},...x_{n}),i=1,2,...,n$.  Now the
non-vanishing components of the Einstein field equation
$$
G_{\mu\nu}=T_{\mu\nu}+\Lambda g_{\mu\nu}
$$

for the above space-time model (1) with matter field in the form
of (2) are [13, 14]

\begin{eqnarray*}
n\dot{\alpha}\dot{\beta}+\frac{1}{2}n(n-1)\dot{\beta}^{2}-e^{-2\beta}\sum_{i=1}^{n}
\left\{\alpha_{x_{i}}^{2}+\frac{1}{2}(n-1)(n-2)\beta_{x_{i}}^{2}+
(n-2)\alpha_{x_{i}}\beta_{x_{i}}+\alpha_{x_{i}x_{i}}\right.
\end{eqnarray*}
\vspace{-8mm}

\begin{equation}
\left.+(n-1)\beta_{x_{i}x_{i}} \right\}+e^{-2\alpha}
\left\{n\alpha'\beta'-\frac{1}{2}n(n+1)\beta'^{2}
-n\beta''\right\}=\Lambda+\rho
\end{equation}

\begin{eqnarray*}
\frac{1}{2}n(n+1)\dot{\beta}^{2}+n\ddot{\beta}-\frac{1}{2}n(n-1)e^{-2\alpha}\beta'^{2}
-e^{-2\beta}\sum_{i=1}^{n}\left\{\frac{1}{2}(n-1)(n-2)\beta_{x_{i}}^{2}+
(n-1)\beta_{x_{i}x_{i}}\right\}
\end{eqnarray*}
\vspace{-8mm}

\begin{equation}
=\Lambda-p \hspace{-3.4in}
\end{equation}

\begin{eqnarray*}
\dot{\alpha}^{2}+\ddot{\alpha}+(n-1)\dot{\alpha}\dot{\beta}+\frac{1}{2}n(n-1)\dot{\beta}^{2}+
(n-1)\ddot{\beta}+e^{-2\alpha}\left\{(n-1)\alpha'\beta'-\frac{1}{2}n(n-1)
\beta'^{2}-\right.
\end{eqnarray*}
\vspace{-8mm}

\begin{eqnarray*}
\left.(n-1)\beta''\right\}-e^{-2\beta}\sum_{i\ne
j=1}^{n}\left\{\alpha_{x_{j}}^{2}+\frac{1}{2}(n-2)(n-3)\beta_{x_{j}}^{2}+
\alpha_{x_{j}x_{j}}+(n-2)\beta_{x_{j}x_{j}}+(n-3)\alpha_{x_{j}}\beta_{x_{j}}\right\}
\end{eqnarray*}
\vspace{-8mm}

\begin{equation}
-e^{-2\beta}\left\{(n-1)\alpha_{x_{i}}\beta_{x_{i}}+
\frac{1}{2}(n-1)(n-2)\beta_{x_{i}}^{2}\right\}=\Lambda-p
\hspace{-.6in}
\end{equation}

\begin{equation}
\alpha_{x_{j}}(-\alpha_{x_{i}}+\beta_{x_{i}})+\beta_{x_{j}}(\alpha_{x_{i}}+
(n-2)\beta_{x_{i}})-\alpha_{x_{i}x_{j}}-(n-2)\beta_{x_{i}x_{j}}=0,~~~
(i\ne j)
\end{equation}

\begin{equation}
(\dot{\beta}-\dot{\alpha})\beta'+\dot{\beta}'=\frac{1}{n}~q~
e^{2\alpha}
\end{equation}

\begin{equation}
(\dot{\alpha}-\dot{\beta})\alpha_{x_{i}}+\dot{\alpha}_{x_{i}}+(n-1)\dot{\beta}_{x_{i}}
=q_{i}~e^{2\beta}
\end{equation}

\begin{equation}
\alpha_{x_{i}}\beta'-\beta'_{x_{i}}=0
\end{equation}

where dot, dash and subscript stands for partial derivatives with
respect to $t,~r$ and the corresponding variables respectively
(e.g. $\beta_{x_{i}}=\frac{\partial\beta}{\partial x_{i}}$) and
$i,j=1,2,3,...,n$.\\

From equations (7) and (9) after differentiating with respect to
$x_{i}$ and $t$ respectively, we have the integrability condition

\begin{equation}
\frac{\partial q}{\partial x_{i}}
+q\alpha_{x_{i}}=n\beta^{'}\dot{\beta}_{x_{i}}e^{-2\alpha},~~~(i=1,2,...,n)
\end{equation}

This equation can not be solved in general. So we have assumed
$\beta'\neq 0, \dot{\beta}_{x_{i}}=0$. Then the form of $\beta$ is

\begin{equation}
e^{\beta}=R(t,r) e^{\nu(r,x_{1},x_{2},...,x_{n})}
\end{equation}

and from equation (9) we have the solution for $\alpha$ as

\begin{equation}
e^{\alpha}=\frac{R'+R\nu'}{D(t,r)}
\end{equation}

where $R$ and $D$ are functions of $t,~r$ only. From equations (4)
and (5) using equations (11) and (12) we have the differential
equations of $R$ and $D$:

\begin{equation}
2R\ddot{R}+(n-1)(\dot{R}^{2}-D^{2})-\frac{2}{n}(\Lambda-p)R^{2}=(n-1)f(r)
\end{equation}
and
\begin{equation}
R\dot{D}=f(r)~ e^{-2\alpha}
\end{equation}

where $f(r)$ is the arbitrary function of $r$.\\

The function $\nu$ satisfies the equation

\begin{equation}
e^{-\nu}=A(r)\sum_{i=1}^{n} x_{i}^{2}+\sum_{i=1}^{n} B_{i}(r)
x_{i}+C(r)
\end{equation}

where $A, ~B_{i}, ~C$ are arbitrary functions of $r$ alone with
the restriction

\begin{equation}
\sum_{i=1}^{n} B_{i}^{2}-4AC=f(r)
\end{equation}

Now from equations (7),(8) and (10) we have the components of heat
flux vector as

\begin{equation}
q=\frac{n}{R}\dot{D}~ e^{-\alpha}
\end{equation}
and
\begin{equation}
q_{i}=-\frac{\dot{D}}{D}\alpha_{x_{i}}e^{-\beta}
\end{equation}

From the above solution, we can see that the field equation (6) is
automatically satisfied. Using (3),(4) and (5) we have the
expression for density as

\begin{equation}
\rho=-\frac{n}{(n-1)}(\ddot{\alpha}+n
\ddot{\beta}+\dot{\alpha}^{2}+n\dot{\beta}^{2})+\frac{(n+1)}{(n-1)}~
p+\frac{2\Lambda}{(n-1)}
\end{equation}

However, from equation (14) we note that as $R$ and $D$ are
functions of $t$ and $r$ only, so $\alpha$ is independent of the
space co-ordinates $x_{i}$~'s ($i=1,2,...,n$) i.e.,
$\alpha_{x_{i}}=0$, $\forall ~i=1,2,...,n$. Hence from equation
(18) we have $q_{i}=0$ and from equation (17) we have seen that
$q=q(t,r)$ i.e., $q$ is a function of $t$ and $r$ only. Thus only
radial heat flow is possible for the choice of the metric as
we consider.\\

\section{\normalsize\bf{Junction conditions and consequence for Szekeres model with
plane symmetric Vaidya metric}}

Let us consider a time-like $(n+1) D$ hypersurface $\Sigma$, which
divides $(n+2)D$ space-time into two distinct $(n+2)D$ manifolds
$V^{-}$ and $V^{+}$. For junction conditions we follow the
modified version of Israel [15] by Santos [4, 5]. Now the geometry
of the space-time $V^{-}$ is given by equation (1) while $V^{+}$
and the boundary $\Sigma$ are cheracterised by the metric ansatzs
as

\begin{equation}
ds_{+}^{2}=\frac{2m(v)}{(n-1)z^{n-1}}dv^{2}-2dvdz+z^{2}
\sum_{i=1}^{n}dx_{i}^{2}
\end{equation}
and
\begin{equation}
ds_{\Sigma}^{2}=-d\tau^{2}+A^{2}(\tau) \sum_{i=1}^{n}dx_{i}^{2}
\end{equation}

where the arbitrary function $m(v)$ in the Vaidya metric
represents the mass at retarded time $v$ inside the boundary
surface $\Sigma$. Now Israel's junction conditions (as described
by Santos) are\\

(i)~ The continuity of the line element i.e.,
\begin{equation}
(ds^{2}_{-})_{\Sigma}=(ds^{2}_{+})_{\Sigma}=ds^{2}_{\Sigma}
\end{equation}

where $(~ )_{\Sigma}$ means the value of (~ ) on $\Sigma$.\\

(ii)~ The continuity of extrinsic curvature over $\Sigma$ gives
\begin{equation}
[K_{ij}]=K_{ij} ^{+}-K_{ij}^{-}=0~,
\end{equation}
where due to Eisenhart the extrinsic curvature has the expresion

\begin{equation}
K_{ij}^{\pm}=-n_{\sigma}^{\pm}\frac{\partial^{2}\chi^{\sigma}_{\pm}}{\partial\xi^{i}
\partial \xi^{j} }-n^{\pm}_{\sigma}\Gamma^{n}_{\mu
\nu}\frac{\partial\chi^{\mu}_{\pm}}{\partial\xi^{i}}\frac{\partial\xi^{\nu}_{\pm}}
{\partial\xi^{j}}
\end{equation}
Here $ \xi^{i}=(\tau,x_{1},x_{2},...,x_{n})$ are the intrinsic
co-ordinates to $\Sigma, ~\chi^{\sigma}_{\pm},~ \sigma
=0,1,2,...,n+1$ are the co-ordinates in $V^{\pm}$ and
$n_{\alpha}^{\pm}$ are the components of the normal vector to
$\Sigma$ in the co-ordinates $\chi^{\sigma}_{\pm}$. It is to be
noted that the above continuity conditions are equivalent to
junction conditions due to Lichnerowicz and O'~Brien and synge.\\

Now for the interior space-time described by the metric (1) the
boundary of the interior matter distribution (i.e., the surface
$\Sigma$) will be characterized by
\begin{equation}
f(r,t)=r-r_{_{\Sigma}}=0
\end{equation}
where $r_{_{\Sigma}}$ is a constant. As the vector with components
$\frac{\partial f}{\partial \chi^{\sigma}_{-} }$ is orthogonal to
$\Sigma$ so we take
$$n_{\mu}^{-}=(0,e^{\alpha},0,...,0).$$
So comparing the metric ansatzs given by equations (1) and (21)
for $dr=0$ we have from the continuity relation (22)

\begin{equation}
\frac{dt}{d\tau}=1,~~A(\tau)=e^{\beta}~~~~
\text{on}~~~r=r_{_{\Sigma}}
\end{equation}

Also the components of the extrinsic curvature for the interior
space-time are

\begin{equation}
K^{-}_{\tau\tau}=0~~~\text{and}~~~  K
^{-}_{x_{i}x_{i}}=\left[\beta'
e^{2\beta-\alpha}\right]_{\Sigma},~~i=1,2,...,n.
\end{equation}

On the other hand for the exterior Vaidya metric described by the
equation (20) with its exterior boundary, given by
\begin{equation}
f(z,v)=z-z_{_{\Sigma}}(v) =0
\end{equation}

the unit normal vector to $\Sigma$ is given by

\begin{equation}
n_{\mu}^{+}=\left(-\frac{2m(v)}{(n-1)z^{n-1}}+2\frac{dz}{dv}\right)^{-1/2}\left(-
\frac{dz}{dv},1,0,...,0\right)
\end{equation}

and the components of the extrinsic curvature are

\begin{equation}
K^{+}_{\tau\tau}=\left[\frac{d^{2}v}{d\tau^{2}}\left(\frac{dv}{d\tau}\right)^{-1}
-\frac{m}{z^{n}}\frac{dv}{d\tau}\right]_{\Sigma}
\end{equation}
and
\begin{equation}
K^{+}_{x_{i}x_{i}}=\left[z~\frac{dz}{d\tau}-\frac{2m(v)}{(n-1)z^{n-2}}\frac{dv}{d\tau}
\right]_{\Sigma},~~~i=1,2,...,n
\end{equation}

Hence the continuity of the extrinsic curvature due to junction
condition (see eq. (23)) gives

\begin{equation}
\left[\frac{2m}{(n-1)}~e^{-(n-2)\beta}~\frac{dv}{d\tau}\right]_{\Sigma}=
\left[e^{2\beta}\left(\dot{\beta}-\beta'e^{-\alpha}\right)\right]_{\Sigma}
\end{equation}
and
\begin{equation}
\left[\frac{1}{\frac{dv}{d\tau}}\right]_{\Sigma}=\left[e^{\beta}\left(\dot{\beta}+
\beta'e^{-\alpha}\right)\right]_{\Sigma}
\end{equation}

Now using the junction condition (32) with the help of equation
(33), we have

\begin{equation}
m(v)=\frac{1}{2}\left[(n-1)e^{(n+1)\beta}\left(\dot{\beta}^{2}-
\beta'^{2}~e^{-2\alpha}\right)\right]_{\Sigma}
\end{equation}

We can interprete this as the total energy bounded within the
surface $\Sigma$ and is equivalent to the well known mass
function in spherical symmetry due to Cahill and McVittie [16].\\

Further using (7), (11), (15), (16), (32), (33) and (34) we
obtain (after simplification)

\begin{equation}
p_{_{\Sigma}}=\left(qe^{\alpha}\right)_{\Sigma}+\frac{1}{2}~n(n-1)
\left[f(r)R^{-2}\right]_{\Sigma}
\end{equation}

where the second term on the r.h.s. does not vanish on $\Sigma$.
So on the boundary vanishing of the isotropic pressure does not
imply the vanishing of the heat flux. Thus for a quasi-spherical
shearing distribution of a collapsing fluid, undergoing
dissipation in the form of heat flow, the isotropic pressure on
the surface of discontinuity $\Sigma$ does not balance the
radiation. Hence in the absence of isotropic pressure there may
still be radiation on the boundary and the exterior space-time
$V^{+}$ will still be Vaidya space-time.\\

Moreover, the total luminosity for an observer at rest at
infinity is

\begin{eqnarray*}
\begin{array}{c}L_{\infty}=
lim~~~ \left(\frac{n-1}{n}\right)z^{n}\epsilon\\
r\rightarrow 0\\
\end{array}
\begin{array}{c}
=-\left(\frac{dm}{dv}\right)_{\Sigma}\\
{}
\end{array}
\end{eqnarray*}
\vspace{-5mm}

\begin{equation}
=\left[\left(\frac{n-1}{n}\right)e^{(n+2)\beta}\frac{(\dot{R}+D)^{2}}{R^{2}}\left\{p-
\frac{1}{2}n(n-1)f(r)R^{-2}\right\} \right]_{\Sigma}
\end{equation}

If we now consider an observer on the boundary $\Sigma$ then the
luminosity for that observer is

\begin{equation}
L_{\Sigma}=\left(\frac{n-1}{n}\right)z^{n}\epsilon_{_{\Sigma}}
=-\left[\left(\frac{dv}{d\tau}\right)^{2}\frac{dm}{dv}\right]_{\Sigma}
\end{equation}
\vspace{-5mm}

\begin{equation}
=\left[\left(\frac{n-1}{n}\right)e^{n\beta}\left\{p-
\frac{1}{2}n(n-1)f(r)R^{-2}\right\} \right]_{\Sigma}
\end{equation}

Thus the boundary red-shift ($Z_{\Sigma}$) of the radiation
emitted by a star can be written as

\begin{equation}
Z_{\Sigma}=\sqrt{\frac{L_{\Sigma}}{L_{\infty}}}~~-1=\left[\frac{e^{-\beta}R}{(\dot{R}+D)}
\right]_{\Sigma}~~-1
\end{equation}

Hence the luminosity measured by an observer at rest at infinity
is reduced by the red-shift in comparison to the luminosity
observed on the surface of the collapsing body. Also when
$\dot{R}+D=0$ then the boundary red-shift attains unlimited value
(i.e., $Z_{\Sigma}\rightarrow\infty$) and the luminosity vanishes
at infinity (i.e., $L_{\infty}\rightarrow 0$).\\

We now discuss the thermodynamical relations for a collapsing
star. We have seen above that when the star produces unpolarized
radiation while its non-adiabatic fluid collapses then the
junction condition (35) between pressure $p$ and heat flux $q$
has to be justified. Also for physically reasonable fluid we
should have (i)~ $0<p<\rho<\infty$ for $0\le r\le r_{_{\Sigma}}$,
(ii)~ both $\frac{d\rho}{d r}$ and $\frac{dp}{dr}$ are negative
for $r>0$, while $0<\frac{dp}{d\rho}<1$ for $0\le r\le
r_{_{\Sigma}}$. Further from thermodynamical point of view, we
should have the following relations [17]:\\

$(a)$~~ $\left(\rho_{_{E}}u^{\mu}\right)_{;\mu}=0$~~~(equation of
conservation of matter) where the effective rest mass density
(measured in the rest frame of $u^{\mu}$), $\rho_{_{E}}$ is
related to the internal energy density $U$ by the relation
$$
\rho=\rho_{_{E}}(1+U)
$$

$(b)$~~ Gibbs equation:
$$
TdS=dU+p~d(1/\rho_{_{E}})
$$
where as usual $S$ is the entropy and $T$ is the temperature [18]
of
the collapsing star.\\

$(c)$~~ Second law of thermodynamics:
$$
S^{\mu}_{;\mu}\ge 0
$$
where the entropy flux $S^{\mu}$ is defined by
$$
S^{\mu}=\rho_{_{E}}S~u^{\mu}+\frac{1}{T}~q^{\mu}
$$
$(d)$~~ Temperature gradient law:
$$
q^{\mu}=-\kappa(g^{\mu\nu}+u^{\mu}u^{\nu})(T,_{\nu}+Tu_{\nu;a}u^{a})
$$
with positive thermal conductivity $\kappa$. But we note that the
second law of thermodynamics can be derived from the temperature
gradient law, so there is no need to satisfy it.\\

Now from the conservation of mass density we have for the present
model
$$
\rho_{_{E}}=\rho_{_{0}}(r,x_{1},...,x_{n})~e^{-(\alpha+n\beta)}
$$
with $\rho_{_{0}}$ as effective rest mass density in the infinite
past.\\

Also from the equation for temperature gradient law the radial
heat flow has the form

\begin{equation}
q=-\kappa~\frac{\partial T}{\partial r}~e^{-2\alpha}
\end{equation}

Further, if we assume the thermal conductivity $\kappa$ as a
polynomial in temperature i.e.,
$$
\kappa=\gamma ~T^{\Omega}\ge 0
$$
then from the comparative study of the expressions for $q$ in
equations (17) and (39) we have the expression for temperature

\begin{equation}
T^{\Omega+1}=-\frac{n(\Omega+1)}{\gamma}\int\left(\frac{R'}{R}+\nu'\right)
\frac{\partial}{\partial t}(log~D)dr+T_{0}(t)
\end{equation}

with $T_{0}(t)$ an arbitrary function of $t$.\\

\section{\normalsize\bf{Discussion}}
In this work we have found a general solution for Szekeres'
$(n+2)$-D space-time model with perfect fluid and heat flux. Here
we may mention that this solution is not only a higher
dimensional generalization of Goode [19] but also it is general
in the sense that here heat flux is directed along all spatial
directions. However, for the study of junction conditions with
exterior Vaidya model we have considered only radial heat flow. We
have studied both the physical conditions and the thermodynamical
relations for the collapse of a radiating star
model.\\

Moreover, in the Szekeres' model if we consider the co-ordinate
transformation [13, 14] $(x_{1},x_{2},...,x_{n})\longrightarrow
(\theta_{1},\theta_{2},...,\theta_{n})$ by

\begin{eqnarray}\begin{array}{llll}
x_{1}=Sin\theta_{n}Sin\theta_{n-1}...~~ ...
Sin\theta_{2}Cot\frac{1}{2}\theta_{1}\\\\
x_{2}=Cos\theta_{n}Sin\theta_{n-1}...~~ ...
Sin\theta_{2}Cot\frac{1}{2}\theta_{1}\\\\
x_{3}=Cos\theta_{n-1}Sin\theta_{n-2}...~~ ...Sin\theta_{2}Cot\frac{1}{2}\theta_{1}\\\\
....~~ ...~~ ...~~ ...~~ ...~~ ...~~ ...~~ ...\\\\
x_{n-1}=Cos\theta_{3}Sin\theta_{2}Cot\frac{1}{2}\theta_{1}\\\\
x_{n}=Cos\theta_{2}Cot\frac{1}{2}\theta_{1}
\end{array}\nonumber
\end{eqnarray}

then the form of the $(n+2)$-D metric equation (1) becomes

\begin{equation}
ds_{-}^{2}=-dt^{2}+e^{2\alpha}dr^{2}+\frac{1}{4}~e^{2\beta}cosec^{4}(\theta_{1}/2)
(d\theta_{1}^{2}+sin^{2}\theta_{1}d\theta_{2}^{2}+.........+
sin^{2}\theta_{1}...sin^{2}\theta_{n-1}d\theta_{n}^{2})
\end{equation}

Now if we match this interior metric with the spherical $(n+2)$-D
Vaidya metric i.e.,

\begin{equation}
ds_{+}^{2}=-\left(1-\frac{2m(v)}{(n-1)z^{n-1}}\right)dv^{2}-2dvdz+z^{2}
(d\theta_{1}^{2}+sin^{2}\theta_{1}d\theta_{2}^{2}+.........+
sin^{2}\theta_{1}...sin^{2}\theta_{n-1}d\theta_{n}^{2})
\end{equation}

then on the boundary the relation between pressure and heat flux
is
$$
p_{_{\Sigma}}=\left(qe^{\alpha}\right)_{\Sigma}~~,
$$
which is identical in form to the earlier works of Santos et al
[4, 5] and Ghosh et al [11]. Also the other conclusions are very
much similar
to their results. So we have not mentioned them here.\\

Therefore, for quasi-spherical radiating star with plane
symmetric boundary the shear-free distribution of collapsing
fluid undergoing dissipation in the form of heat flux, it is not
necessary to have non-vanishing isotropic pressure on the
boundary for existence of heat flow on it and the result strongly
depends on the value of $f(r)$.\\

{\bf Acknowledgement:}\\\\
U.D. is thankful to C.S.I.R., Govt. of  India  for  awarding  a  Junior  Research
Fellowship.\\

{\bf References:}\\
\\
$[1]$  J. R. Oppenhiemer and H. Snyder, {\it Phys. Rev.} {\bf 56} 455 (1939).\\
$[2]$  C. W. Misner and D. Sharp, {\it Phys. Rev.} {\bf 136} b571
(1964).\\
$[3]$  P. C. Vaidya, {\it Proc. Indian Acad. Sci. A} {\bf 33} 264
(1951).\\
$[4]$  N. O. Santos, {\it Phys. Lett. A} {\bf 106} 296 (1984).\\
$[5]$  N. O. Santos, {\it Mon. Not. R. Astr. Soc.} {\bf 216} 403
(1985).\\
$[6]$  A. K. G. de Oliveira, N. O. Santos and C. A. Kolassis,
{\it Mon. Not. R. Astr. Soc.} {\bf 216} 1001
(1985).\\
$[7]$  A. K. G. de Oliveira, J. A. de F. Pacheco and N. O. Santos,
{\it Mon. Not. R. Astr. Soc.} {\bf 220} 405
(1986).\\
$[8]$  A. K. G. de Oliveira and N. O. Santos, {\it Astrophys. J.}
{\bf 312} 640 (1987).\\
$[9]$  A. K. G. de Oliveira, C. A. Kolassis and N. O. Santos, {\it
Mon. Not. R. Astr. Soc.} {\bf 231} 1011 (1988).\\
$[10]$  S. G. Ghosh and D. W. Deshkar, {\it Int. J. Mod. Phys. D}
{\bf 12} 317 (2003).\\
$[11]$  S. G. Ghosh and D. W. Deshkar, {\it Gravitation and
Cosmology} {\bf 6} 1 (2000).\\
$[12]$  P. Szekeres, {\it Commun. Math. Phys.} {41} 55 (1975).\\
$[13]$  S. Chakraborty and U. Debnath, {\it gr-qc}/0304072.\\
$[14]$  U. Debnath, S. Chakraborty and J. D. Barrow,
{\it gr-qc}/0305075 (Accepted in Gen. Rel. Grav., 2003).\\
$[15]$  W. Israel, {\it Nuovo Cimento} {\bf 44B} 1 (1966).\\
$[16]$  M. E. Cahill and G. C. McVittie, {\it J. Math. Phys.} {\bf
11} 1382 (1970).\\
$[17]$  R. Treciokas and G. F. Ellis, {\it Commun. Math. Phys.}
{\bf 23} 1 (1971).\\
$[18]$  C. Eckart, {\it Phys. Rev.} {\bf 58} 919 (1940).\\
$[19]$  S. W. Goode, {\it Class. Quantum Grav.} {\bf 3} 1247
(1986).\\

\end{document}